\documentclass[conference]{IEEEtran}
\IEEEoverridecommandlockouts
\usepackage{cite}
\usepackage{amsmath,amssymb,amsfonts}
\usepackage{algorithmic}
\usepackage{graphicx}
\usepackage{textcomp}
\usepackage{hyperref}
\usepackage{xcolor}
\usepackage{amsmath}
\def\BibTeX{{\rm B\kern-.05em{\sc i\kern-.025em b}\kern-.08em
    T\kern-.1667em\lower.7ex\hbox{E}\kern-.125emX}}
\begin{document}

\title{Seasonality Patterns in 311-Reported Foodborne Illness Cases and Machine Learning-Identified Indications of Foodborne Illnesses from Yelp Reviews, New York City, 2022-2023\\

}

\author{\IEEEauthorblockN{Eden Shaveet}
\IEEEauthorblockA{\textit{Dept. Computer Science} \\
\textit{Columbia University}\\
New York, USA \\
ems2349@columbia.edu}
\and
\IEEEauthorblockN{Crystal Su}
\IEEEauthorblockA{\textit{Dept. Computer Science} \\
\textit{Columbia University}\\
New York, USA \\
ys3791@columbia.edu}
\and
\IEEEauthorblockN{Daniel Hsu}
\IEEEauthorblockA{\textit{Dept. Computer Science} \\
\textit{Columbia University}\\
New York, USA \\
djhsu@cs.columbia.edu}
\and
\IEEEauthorblockN{Luis Gravano}
\IEEEauthorblockA{\textit{Dept. Computer Science} \\
\textit{Columbia University}\\
New York, USA \\
gravano@cs.columbia.edu}
}

\maketitle

\begin{abstract}
Restaurants are critical venues at which to investigate foodborne illness outbreaks due to shared sourcing, preparation, and distribution of foods. Formal channels to report illness after food consumption, such as 311, New York City’s non-emergency municipal service platform, are underutilized. Given this, online social media platforms serve as abundant sources of user-generated content that provide critical insights into the needs of individuals and populations. We extracted restaurant reviews and metadata from Yelp to identify potential outbreaks of foodborne illness in connection with consuming food from restaurants. Because the prevalence of foodborne illnesses may increase in warmer months as higher temperatures breed more favorable conditions for bacterial growth, we aimed to identify seasonal patterns in foodborne illness reports from 311 and identify seasonal patterns of foodborne illness from Yelp reviews for New York City restaurants using a Hierarchical Sigmoid Attention Network (HSAN). We found no evidence of significant bivariate associations between any variables of interest. Given the inherent limitations of relying solely on user-generated data for public health insights, it is imperative to complement these sources with other data streams and insights from subject matter experts. Future investigations should involve conducting these analyses at more granular spatial and temporal scales to explore the presence of such differences or associations.
\end{abstract}

\section{Introduction}
\subsection{Foodborne Illness \& Infodemiology}
Foodborne illnesses are gastrointestinal conditions traced to consuming food contaminated by hazardous bacteria, toxins, parasites, viruses, or chemicals \cite{b1}. Restaurants are critical venues at which to investigate foodborne illness outbreaks due to shared sourcing, preparation, and distribution of foods. According to the national Foodborne Disease Outbreak Surveillance System (FDOSS), more than 60\% of reported foodborne illness outbreaks between 2016 and 2018 were believed to originate at a restaurant or food service establishment \cite{b2}. As social media platforms have gained widespread popularity over the past two decades \cite{b3}, people and communities have increasingly turned to these platforms as a means of sharing health-related incidents as opposed to formal reporting channels \cite{b4} \cite{b5}. Formal channels to report illness after food consumption, such as 311, New York City’s non-emergency municipal service platform, are underutilized \cite{b6}. Given this, online social media platforms serve as abundant sources of user-generated content that provide critical insights into the needs of individuals and populations \cite{b7}. The extraction of user-generated information from online sources, including social media platforms, to gain public health insights via social listening methods is central to an emerging field of study known as infodemiology. Infodemiological methods focus on scanning the internet for user-contributed health-related content, with the ultimate goal of informing and improving public health interventions \cite{b8}.

The Adaptive Information Extraction from Social Media for Actionable Inferences in Public Health project is a collaboration between the Columbia University Department of Computer Science and the New York City Department of Health and Mental Hygiene (NYC DOHMH) in which we extract restaurant reviews and metadata from Yelp to identify potential outbreaks of foodborne illness in connection with consuming food from restaurants. Each day, we retrieve data from Yelp inc. and process free-text reviews via a Hierarchical Sigmoid Attention Network (HSAN) classifier to indicate the likelihood that a review is related to a foodborne illness event \cite{b9}.

\subsection{Foodborne Illness Seasonality}
Between 1998 and 2013, improper food handling and preparation practices, such as incorrect temperature control practices and cross-contamination of foods, were primary contributors to restaurant-associated illness outbreaks \cite{b10}. Norovirus was the most common restaurant-associated foodborne illness outbreak during this time, accounting for 46\% of all reported outbreaks \cite{b10}. While much of this can be attributed to mishandling of food and ingredients by food service staff, an additional consideration is how outdoor temperature and weather patterns impact food storage conditions and practices. For example, outdoor conditions can influence the indoor environment of food preparation establishments or the conditions within delivery vehicles, altering the requirements for effective temperature control and food handling practices \cite{b11}. Additionally, there is concern around how climate can affect food safety and spoilage at every point in the food chain \cite{b12}. It has been speculated that the prevalence of foodborne illnesses may increase in warmer months as higher temperatures breed more favorable conditions for bacterial growth \cite{b13}. Between 1996 and 2017, seasonal peaks of foodborne illnesses including Campylobacter, Salmonella, and Shiga toxin-producing Escherichia Coli (STEC) were observed in late-July at both national and state levels \cite{b14}. While seasonality patterns are well-established in food safety literature, seasonality of foodborne illness has yet to be explored in infodemiological contexts.

\subsection{Objectives}
The objectives of our study were to 1.) identify seasonal patterns in foodborne illness reports from 311, New York City’s non-emergency municipal service platform on which residents can report issues, including restaurant-associated foodborne illness \cite{b15} and 2.) identify seasonal patterns of foodborne illness from Yelp reviews for New York City Restaurants using a Hierarchical Sigmoid Attention Network (HSAN) to assign classification scores that indicate the likelihood that a review is describing a foodborne illness event.

\section{Methods}
We extracted reviews and business addresses via a private application programming interface (API) between Yelp Inc. and Columbia University. We scored reviews using an HSAN classifier, which 1.) encodes each segment of the review using word embeddings and convolutional neural networks (CNNs), 2.) classifies each segment using a softmax classifier, and 3.) aggregates the segment predictions to compute an overall review prediction. HSAN produces a score between 0-1, where higher scores indicate higher likelihood that a review is describing a foodborne illness event. For more information about HSAN, see Karamanolakis et al., 2019. This analysis was limited to Yelp reviews for whom “non-zero” HSAN scores were assigned (scored \textgreater 0.05).

To investigate the seasonality of free-text indications of foodborne illness on Yelp, we assessed the correlation between average daily temperature and average daily HSAN score in 2022-2023 by Spearman’s rank correlation coefficient as well as the distribution of average daily HSAN scores in NYC by season (Winter: December, January, February; Spring: March, April, May; Summer: June, July, August; Fall: September, October, November) in 2022-2023 using a Kruskal-Wallis test. We conducted another Kruskal-Wallis test to assess the distribution of average daily temperatures in NYC by daily frequency of Yelp reviews likely to be indicative of foodborne illness (HSAN score \textgreater 0.5).

To investigate the seasonality of 311-reported foodborne illness cases, we assessed the distribution of average daily temperature (Fahrenheit) in NYC by the number of 311-reported foodborne illness cases in 2022-2023 using a Kruskal-Wallis test.

\section{Results}
We found no evidence of significant bivariate associations between any variables of interest. Briefly, we found a very weak, non-significant negative association between average daily temperature and average daily HSAN score in 2022-2023 by Spearman’s rank correlation coefficient $(\rho \approx -0.05, p \approx 0.15)$. 
\begin{figure}[htbp]
    \centering
    \includegraphics[width=0.5\textwidth]{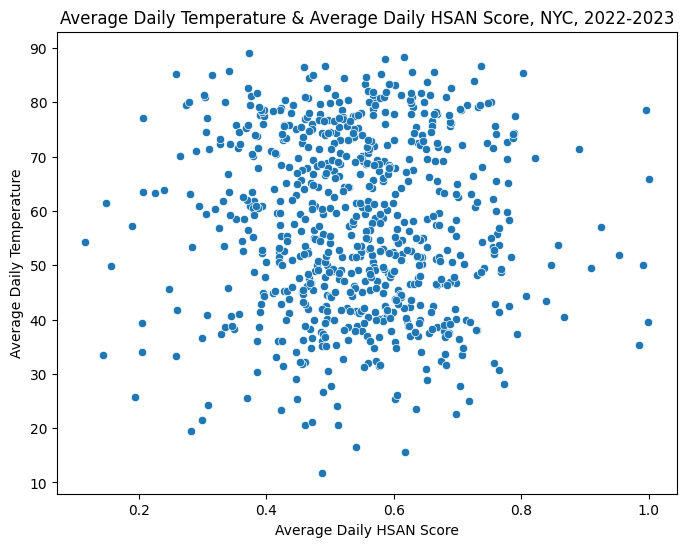}
    \caption{Average Daily Temperature \& Average Daily HSAN Score, NYC, 2022-2023.}
    \label{fig}
\end{figure}

We found no significant difference in the distribution of average daily HSAN scores in NYC by season in 2022-2023 $(H \approx 7.37, p \approx 0.06)$.
\begin{figure}[htbp]
    \centering
    \includegraphics[width=0.5\textwidth]{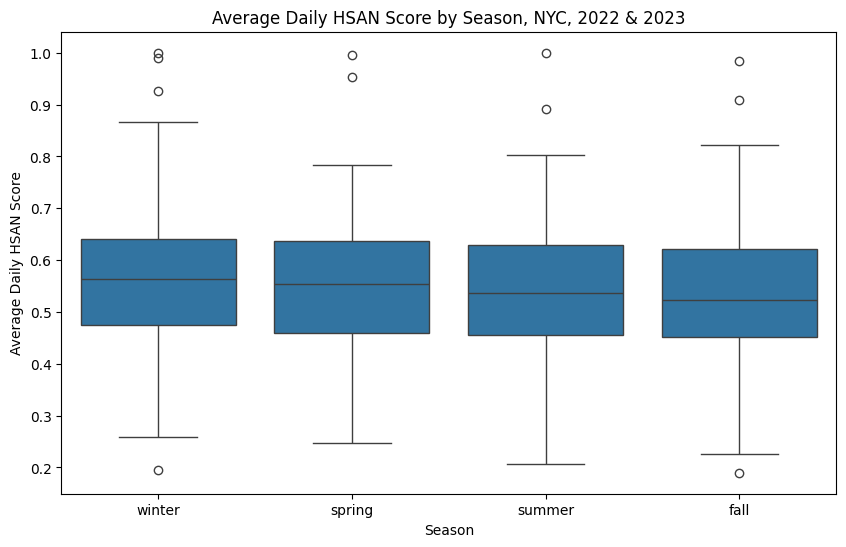}
    \caption{Average Daily HSAN Score by Season, NYC, 2022 \& 2023.}
    \label{fig}
\end{figure}

We found no significant difference in the  distribution of average daily temperatures in NYC by daily frequency of Yelp reviews likely to be indicative of foodborne illness (HSAN score \textgreater 0.5) in 2022-2023 $(H \approx 3.89, p \approx 0.57)$. 
\begin{figure}[htbp]
    \centering
    \includegraphics[width=0.5\textwidth]{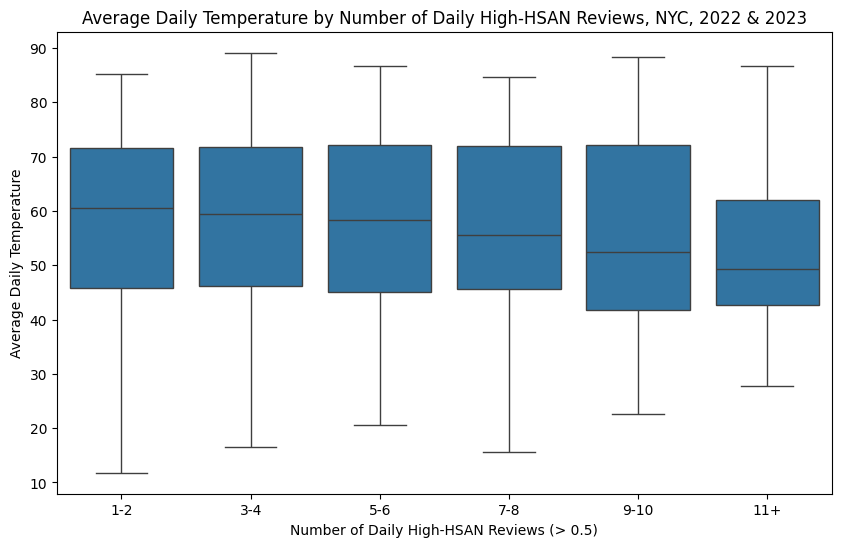}
    \caption{Average Daily Temperature by Number of Daily High-HSAN Reviews, NYC, 2022 \& 2023.}
    \label{fig}
\end{figure}

We found no significant difference in the distribution of average daily temperatures in NYC by the number of 311-reported foodborne illness cases by day in 2022-2023 $(H \approx 3.16, p \approx 0.68)$.
\begin{figure}[htbp]
    \centering
    \includegraphics[width=0.5\textwidth]{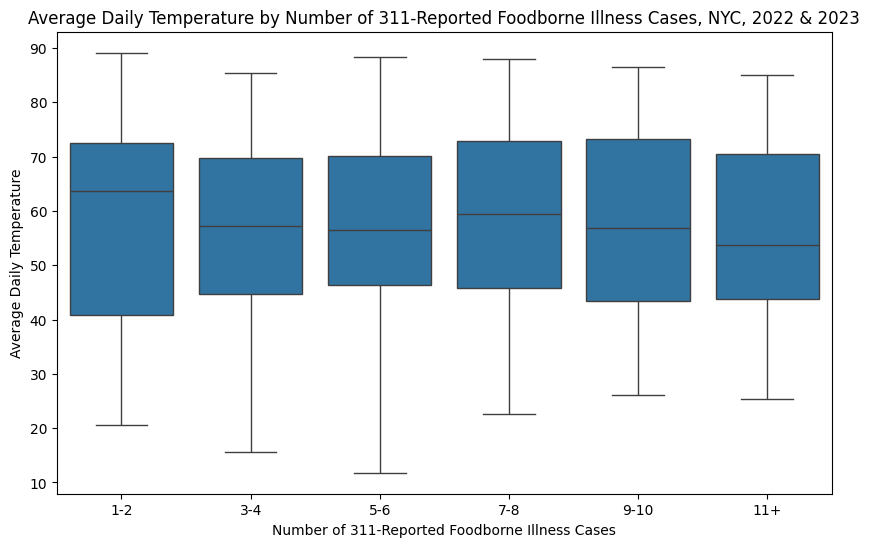}
    \caption{Average Daily Temperature by Number of 311-Reported Foodborne Illness Cases, NYC, 2022 \& 2023.}
    \label{fig}
\end{figure}

\section{Discussion}
While our bivariate analyses did not reveal a significant correlation between seasonality and self-reported or passively detected foodborne illness onset via Yelp reviews at the city level, these findings underscore a critical challenge in infodemiological research. Specifically, how do we reconcile discrepancies between infodemiological observations and established epidemiological knowledge?

The absence of a bivariate correlation between high outdoor temperatures, a known factor associated with foodborne illness outbreaks, and the number of self-reported foodborne illness cases on 311 or language used in Yelp reviews prompts consideration of several possibilities. Perhaps there was truly no difference in foodborne illness onset between days of higher and lower outdoor temperatures in 2022 and 2023. Perhaps cases of foodborne illness were underreported through the 311 channel and underdetected on the Yelp platform. Perhaps finer spatial or temporal resolutions are needed to reveal detectable differences or associations. Or perhaps there is a mediating factor related to Yelp user behavior or the self report process.

It’s important to acknowledge that the onset of foodborne illness is influenced by numerous factors not explored in this study, which could have a more significant impact. For instance, in addition to reasons previously cited, staff practices may vary throughout the year as some restaurants employ seasonal workers with less experience \cite{b16}. Additionally, while changes in outdoor temperatures could necessitate adjustments in indoor food safety protocols, some establishments implement stringent measures to ensure their food and ingredients maintain appropriate temperatures throughout their lifecycle before preparation, mitigating the impact of outdoor temperature fluctuations on food served \cite{b17}. Finally, these analyses were conducted at the city level, as opposed to a more granular spatial unit. New York City is large (1,214 km²) with variable weather patterns across its five boroughs. Given this, future investigations could involve conducting these analyses at more granular spatial and temporal scales to explore the presence of such differences or associations. 

Given the inherent limitations of relying solely on user-generated data for public health insights, it is imperative to complement these sources with other data streams and insights from subject matter experts.

\section*{Acknowledgments}
We thank the New York City Department of Health and Mental Hygiene and Yelp inc. for their ongoing collaboration. The Adaptive Information Extraction from Social Media for Actionable Inferences in Public Health project was supported by the National Science Foundation under Grant No. IIS-15-63785. Any opinions, findings, and conclusions or recommendations expressed in this material are those of the authors and do not necessarily reflect the views of the National Science Foundation.


\begin{thebibliography}{00}
\bibitem{b1} D. C. Pigott, “Foodborne Illness,” Emergency Medicine Clinics of North America, vol. 26, no. 2, pp. 475–497, May 2008, doi: 10.1016/j.emc.2008.01.009.
\bibitem{b2} T. N. Kim et al., “Foodborne Outbreak Rates Associated with Restaurant Inspection Grading and Posting at the Point of Service: Evaluation Using National Foodborne Outbreak Surveillance Data,” Journal of Food Protection, vol. 85, no. 7, pp. 1000–1007, Jul. 2022, doi: 10.4315/JFP-22-007.
\bibitem{b3} Pew Research Center, “Social Media Fact Sheet,” Pew Research Center: Internet, Science \& Tech. Accessed: Mar. 07, 2023. [Online]. Available: https://www.pewresearch.org/internet/fact-sheet/social-media/
\bibitem{b4} W. Ahmed, R. Jagsi, T. G. Gutheil, and M. S. Katz, “Public Disclosure on Social Media of Identifiable Patient Information by Health Professionals: Content Analysis of Twitter Data,” J Med Internet Res, vol. 22, no. 9, p. e19746, Sep. 2020, doi: 10.2196/19746.
\bibitem{b5} S. H. Berg et al., “Health authorities’ health risk communication with the public during pandemics: a rapid scoping review,” BMC Public Health, vol. 21, no. 1, p. 1401, Jul. 2021, doi: 10.1186/s12889-021-11468-3.
\bibitem{b6} T. Effland et al., “Discovering foodborne illness in online restaurant reviews,” Journal of the American Medical Informatics Association, vol. 25, no. 12, pp. 1586–1592, Dec. 2018, doi: 10.1093/jamia/ocx093.
\bibitem{b7} W. Zhuang, Q. Zeng, Y. Zhang, C. Liu, and W. Fan, “What makes user-generated content more helpful on social media platforms? Insights from creator interactivity perspective,” Information Processing \& Management, vol. 60, no. 2, p. 103201, Mar. 2023, doi: 10.1016/j.ipm.2022.103201.
\bibitem{b8} G. Eysenbach, “Infodemiology and Infoveillance: Framework for an Emerging Set of Public Health Informatics Methods to Analyze Search, Communication and Publication Behavior on the Internet,” J Med Internet Res, vol. 11, no. 1, p. e11, Mar. 2009, doi: 10.2196/jmir.1157.
\bibitem{b9} G. Karamanolakis, D. Hsu, and L. Gravano, “Weakly Supervised Attention Networks for Fine-Grained Opinion Mining and Public Health,” in Proceedings of the 5th Workshop on Noisy User-generated Text (W-NUT 2019), Hong Kong, China: Association for Computational Linguistics, 2019, pp. 1–10. doi: 10.18653/v1/D19-5501.
\bibitem{b10} K. M. Angelo, A. L. Nisler, A. J. Hall, L. G. Brown, and L. H. Gould, “Epidemiology of restaurant-associated foodborne disease outbreaks, United States, 1998–2013,” Epidemiology \& Infection, vol. 145, no. 3, pp. 523–534, Feb. 2017, doi: 10.1017/S0950268816002314.
\bibitem{b11} “FSIS Food Safety and Security Guidelines for the Transportation and Distribution of Meat, Poultry, and Egg Products,” United States Department of Agriculture; Food Safety and Inspection Service, 2003.
\bibitem{b12} O. Misiou and K. Koutsoumanis, “Climate change and its implications for food safety and spoilage,” Trends in Food Science \& Technology, vol. 126, pp. 142–152, Aug. 2022, doi: 10.1016/j.tifs.2021.03.031.
\bibitem{b13} nyc.gov, “Hot weather and food safety,” Environment \& Health Data Portal. Accessed: Mar. 25, 2024. [Online]. Available: https://a816-dohbesp.nyc.gov/IndicatorPublic/data-stories/food/
\bibitem{b14} R. B. Simpson, B. Zhou, and E. N. Naumova, “Seasonal synchronization of foodborne outbreaks in the United States, 1996–2017,” Sci Rep, vol. 10, no. 1, p. 17500, Oct. 2020, doi: 10.1038/s41598-020-74435-9.
\bibitem{b15} “NYC 311,” nyc.gov. [Online]. Available: https://portal.311.nyc.gov/
\bibitem{b16} D. Reynolds, E. A. Merritt, and A. Gladstein, “Retention Tactics for Seasonal Employers: An Exploratory Study of U.S.-Based Restaurants,” Journal of Hospitality \& Tourism Research, vol. 28, no. 2, pp. 230–241, May 2004, doi: 10.1177/1096348004263104.
\bibitem{b17} J.-S. Lee et al., “Evaluation of Food Safety Performance and Food Storage Condition in Restaurants against Climate Change,” Journal of Food Hygiene and Safety, vol. 29, no. 3, pp. 195–201, 2014, doi: 10.13103/JFHS.2014.29.3.195.
\end{thebibliography}
\end{document}